\documentclass[12pt]{iopart}
\expandafter\let\csname equation*\endcsname\relax
\expandafter\let\csname endequation*\endcsname\relax
\usepackage[intlimits]{amsmath}	
\usepackage{dsfont}
\usepackage[capitalise,sort,noabbrev]{cleveref}
\crefname{equation}{}{}
\usepackage{amssymb}
\usepackage{csquotes}
\usepackage{graphicx}

  \newcommand{\bra}[1]{\left\langle #1 \right|}
  \newcommand{\ket}[1]{\left| #1\right\rangle}
  \newcommand{\braket}[2]{\langle #1| #2 \rangle}
  
  \newcommand{\abs}[1]{\vert #1 \vert}
  \newcommand{\abssq}[1]{\vert #1 \vert^2}
  \newcommand{\idone}{\hat{\mathds{1}}}

\newcommand{\dif}[2]{\frac{\mathrm{d}#1}{\mathrm{d}#2}}
\newcommand{\pdif}[2]{\frac{\partial #1}{\partial #2}}
\newcommand{\hvsigma}{\hat{\boldsymbol{\sigma}}}

\renewcommand{\vec}[1]{\mathbf{#1}}
\DeclareMathOperator\diag{diag}
\DeclareMathOperator\ii{i}
\begin{document}

\title{Estimating the privacy of quantum-random numbers}

\author{Johannes Seiler$^1$, Thomas Strohm$^2$, and Wolfgang P. Schleich$^{1,3,4}$}

\address{$^1$Institut f{\"u}r Quantenphysik \& Center for Integrated Quantum Science and Technology ($\mathrm{IQ^{ST}}$), Universit{\"a}t Ulm, D-89069 Ulm, Germany}
\address{$^2$Corporate Research, Robert Bosch GmbH, D-71272 Renningen, Germany}
\address{$^3$Institute of Quantum Technologies, German Aerospace Center (DLR), S{\"o}flinger Str.~100, D-89077 Ulm, Germany}
\address{$^4$Hagler Institute for Advanced Study, Institute for
	Quantum Science and Engineering (IQSE), and Texas A{\&}M AgriLife
	Research, Texas A{\&}M University, College Station, TX 77843-4242, USA}

\ead{johannes.seiler@uni-ulm.de} 



\begin{abstract}
We analyze the information an attacker can obtain on the numbers generated by a user by measurements on a subsystem of a system consisting of two entangled two-level systems. The attacker and the user make measurements on their respective subsystems, only. Already the knowledge of the density matrix of the subsystem of the user completely determines the upper bound on the information accessible to the attacker. We compare and contrast this information to the appropriate bounds provided by quantum state discrimination.
%
\end{abstract}
\maketitle

\section{Introduction}
Random numbers have wide applications \cite{Herrero-Collantes2017}, ranging from Monte Carlo simulations \cite{Metropolis1949} via lotteries and gambling to classical and quantum cryptography protocols \cite{Menezes1996,Nielsen2001,Gisin2002,scully1997quantum}.
For most of these tasks, the privacy of the generated numbers, that is the condition that the random numbers are neither predictable by any model, nor that an attacker can obtain information that allows him to at least partially predict them, plays a crucial role.

A quantum random number generator (QNRG) offers at least theoretically the possibility to create such unpredictable random numbers \cite{Acin2016,Ma2016}, due to the physical nature of their generation process and the inherent indeterminism of quantum theory.
Typical examples of QRNG implementations are photons on a beam splitter \cite{Jennewein2000}, homodyne measurements of the vacuum \cite{Gabriel2010}, or laser phase noise \cite{Abellan2015}.

However, real life implementations of QRNG usually suffer from imperfections that open the door for an attacker to get at least partial information about the generated numbers. In this article, we employ an elementary two-qubit model for such a non-ideal QRNG to determine how much information an attacker can maximally gain by exploiting the imperfections of a QRNG. 

We emphasize that our model can be easily implemented experimentally. In order to implement our model experimentally, two conditions have to be fulfilled: (i) The control and entanglement of two qubit systems. (ii)  The tomography of both qubits. Fortunately, can be achieved readily. Over the past years, a wide range of experiments controlling and measuring two qubit systems have been demonstrated, ranging from superconducting qubits \cite{Chiorescu2004}, over trapped ions \cite{Turchette1998,Blatt2008} and Rydberg atoms \cite{Raimond2001}, to entangled photons \cite{Kwiat1995}. Tomography has also been demonstrated for different systems \cite{Steffen2006,Cutshall2020}.

\subsection{Formulation of problem}\label{sec:fotp}
\begin{figure}
	\centering 
	\includegraphics[scale=.5]{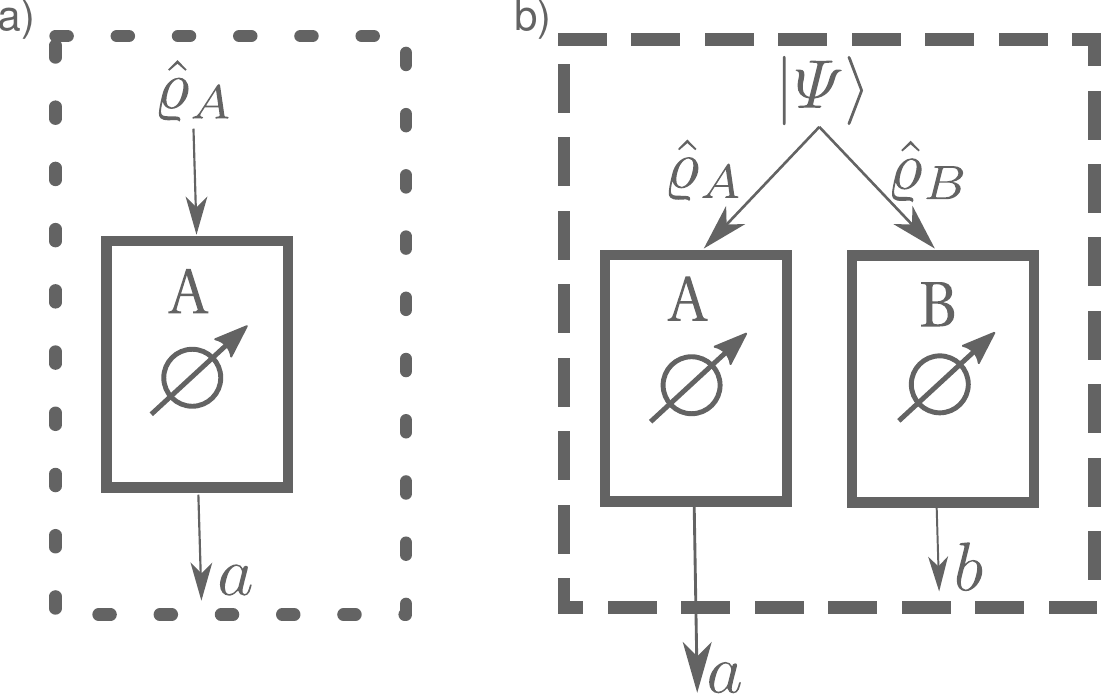}
	\caption{Model of a quantum random number generator based on two entangled qubit systems and viewed from the user a) and the attacker b). a) The user sees a mixed state $\hat{\varrho}_{A}$ and makes a projective measurement yielding a random bit $a$. b) The attacker deals with the complete system $A+B$ in which the mixed state $\hat{\varrho}_{A}$ is purified to $\ket{\Psi}$. The user still performs a measurement on $\hat{\varrho}_{A}$ to obtain the bit $a$, while the attacker carries out a measurement on $\hat{\varrho}_{B}$ to receive a bit $b$. The question is: How much information about $a$ can the attacker obtain from his result $b$?}
	\label{fig:setup}
\end{figure}

For this purpose, we consider the model of a QRNG depicted in \cref{fig:setup} which consists of a single qubit system $A$, that is prepared in a quantum state~$\hat{\varrho}_{A}$. 
The user performs projective measurements in the direction of the unit vector $\vec{e}_{A}$ on the Bloch sphere of the system $A$. 
To each of the two possible outcomes he assigns a bit value $a$, with $ a=0 $ or $ a=1 $. We denote the probability that the user obtains the bit value $a$ for the measurement direction $\vec{e}_{A}$ by $W_{\vec{e}_{A}}(a)$.

Since the user wants to extract a maximum of entropy, his measurement is chosen in a way, that the measurement outcomes, and thus the assigned bit values, have equal probability. In the ideal case, the state $\hat{\varrho}_{A}$ would be a pure state, but due to imperfections it is in general assumed to be a mixed state. By extending the system with a qubit environment $B$, 
we can purify $\hat{\varrho}_{A}$ to a pure state $\ket{\Psi}$ in the system $A+B$. 

In the worst case, an attacker, who wants to gain as much knowledge about the generated random numbers as possible, knows or might even have prepared the complete state $\ket{\Psi}$. The attacker is also aware of the user's measurement, and can perform a projective measurement on the subsystem $ B $. We denote the measurement direction by the unit vector $\vec{e}_{B}$ on the Bloch sphere of the subsystem $B$. This measurement yields a bit value outcome $b$ with probability $W_{\vec{e}_{B}}(b)$, where $ b=0 $ or $ b=1 $. 

The question the user has to ask then is: How much information can the attacker gain from his own measurement result $b$ about the user's random bit $a$?

\subsection{Mutual information and entanglement}
We quantify this information using the mutual information \cite{Nielsen2001,StigStenholm2005,ThomasM.Cover2006}
\begin{equation}
I(\vec{e}_{A},\vec{e}_{B},\ket{\Psi}) = \sum_{a,b=0}^{1} W_{\vec{e}_{A},\vec{e}_{B}}(a,b)\log_{2}\left(\frac{ W_{\vec{e}_{A},\vec{e}_{B}}(a,b)}{ W_{\vec{e}_{A}}(a)W_{\vec{e}_{B}}(b)}\right),
\label{eq:def-mutual_information}
\end{equation}
that a measurement on the system $B$ can provide about the measurement outcome in the system $A$, and vice versa. Here, $W_{\vec{e}_{A},\vec{e}_{B}}(a,b)$ is the joint probability of getting the measurement results $a$ and $b$. 

%
%

We note, that for a separable state $ \ket{\Psi_{s}} $, the measurement results in both subsystems are independent of each other, that is the joint probability is given by the product
\begin{equation}\label{eq:prob-sep-state}
W_{\vec{e}_{A},\vec{e}_{B}}(a,b) = W_{\vec{e}_{A}}(a)W_{\vec{e}_{B}}(b)
\end{equation}
of the marginals 
for all combinations of measurement results $ a $ and $ b $ and the logarithm and hence the mutual information both vanish, that is
\begin{equation}\label{key}
I(\vec{e}_{A},\vec{e}_{B},\ket{\Psi_{s}})=0.
\end{equation}

In order to achieve a non-vanishing mutual information, the two subsystems $ A $ and~$ B $ must be entangled. Indeed, we shall show that the entanglement between the two subsystems plays a crucial role for the mutual information. 

We gain a deeper insight into the role of the entanglement, by noting from \cref{eq:def-mutual_information} that the mutual information depends only on the measurement probabilities, which result from the measurement operators of the user and the attacker as well as from the state of the complete system. 

Since, we want to model a quantum random number generator, the user chooses the measurement such that a uniform distribution arises. The user's measurement is therefore fixed with respect to the state of the subsystem of the user. The mutual information is then only dependent on the measurement of the attacker and the state of the complete system. 

To obtain the maximal mutual information, the attacker has to choose his measurement accordingly. The requirements of a constant distribution for the user and the maximal mutual information for the attacker reduce the number of degrees of freedom and the mutual information can only depend on the entanglement of the two subsystems.

\subsection{Discussion of the literature}
The question raised in this article of how private the random numbers generated in a non-ideal QRNG are, is of course not completely new. There already exist different approaches \cite{Frauchiger2013,Mitchell2015,Pironio2010,Gallego2010,Bowles2014,Brask2017} that allow to estimate the unpredictability of the \enquote{raw} random numbers generated in a non-ideal QRNG. 
All strategies have in common that one tries to find a lower bound to the min-entropy of a long sequence of raw random numbers. This quantity is then used by a randomness extractor to produce a shorter, but unpredictable sequence of \enquote{perfect} random numbers \cite{Bennett1995,Nisan1999,Ma2013}.

One approach is to model the setup and its imperfections, and then calculate the min-entropy from this model \cite{Frauchiger2013,Mitchell2015}. However, in many cases this is quite a difficult task, and one has to make sure that the model is a good description of the experimental implementation. 

Semi-device independent QRNGs \cite{Pironio2010,Gallego2010,Bowles2014,Brask2017}, in which states are prepared and measured in random bases in order to make Bell-like tests on the raw data represent a different approach. Here, the violation of certain (in-)equalities, for example Bell inequalities \cite{Brunner_2014}, of these data then certifies the non-classicality of the physical process, and determines a lower bound on the min-entropy. This procedure has the advantage that one does not need a specific model of the QRNG, while only certain weaker assumptions on the preparation and/or the measurement devices have to be fulfilled. 

Our approach is very much in the line of Ref.~\cite{Frauchiger2013} but much more specific. In comparison to the latter paper, we discuss how much information an attacker can get, and how this information depends on the measured quantum state and the chosen measurements. This approach gives us the possibility to show how the attacker can gain information, and how the user of the QRNG can protect himself against it. 

Another difference of our approach is that we use the \textit{mutual information} as the quantity of interest instead of the min-entropy. However, our results could also be easily formulated in terms of the latter.

\subsection{Outline}
Our article is organized as follows: In \cref{sec:singleprojmeas}, we consider the case of fixed projective measurement directions in both the system and the environment, and derive a general expression for the mutual information. We then focus in \cref{sec:wcs} on the case of a QRNG, where the user selects his measurement in such a way that the bit $a$ is uniformly distributed, and obtain the maximal information any attacker can gain.  Finally, in \cref{sec:conclusionsandoutlook} we conclude by summarizing our results and providing a short outlook.

In order to keep our article self-contained while focused on the essential ideas we have included additional material and extensive calculations in three appendices. In \ref{app:Acorrel} and \ref{app:para-const} we evaluate explicitly the constraints on three parameters that fully define the mutual information. Moreover, we dedicate \ref{app:maxI} to a detailed derivation of the maximal mutual information. \ref{sec:rm} is devoted to extending the user's measurement strategy.

\section{Mutual information for projective measurements} \label{sec:singleprojmeas}
In this section we derive a general expression for the mutual information in our QRNG model for the case, when only projective measurements are performed on both $A$ and~$B$. We discuss the dependence of the mutual information on the entanglement of the two qubit subsystems as well as on the measurement directions.
The results provided in this section will serve as the foundation of our analysis of the worst case presented in \cref{sec:wcs}.

\subsection{States of system and subsystems}
We start from the pure two-qubit state
\begin{equation}
\ket{\Psi} \equiv \sum_{i=0}^{1}\sum_{j=0}^{1}\Psi_{ij}\ket{i}_{A}\ket{j}_{B},
\label{eq:state-01basis}
\end{equation}
representing the state of the combined system of $A$ and $B$ by complex coefficients~$\Psi_{ij}$, which can be interpreted as the elements of a $ 2\times2 $ matrix $ \Psi $. We quantify the entanglement between the two subsystems of the state $ \ket{\Psi} $ by the concurrence
\begin{equation}\label{eq:def-concurrence0}
\mathcal{C} \equiv 2 \abs{\det\Psi},
\end{equation}
which can take values between zero, for $ \ket{\Psi} $ being a separable state, and one, when $ \ket{\Psi} $ is a maximally entangled state.

When we trace out the subsystem $ B(A) $, we obtain the reduced density operator
\begin{equation}\label{eq:def-rhoaorb}
\hat{\varrho}_{A(B)} \equiv \tr_{B(A)}\left(\ket{\Psi}\bra{\Psi}\right)
\end{equation}
of the subsystem $ A(B) $, which can be written in the form
\begin{equation}\label{eq:def-rhoaorb-blochv}
\hat{\varrho}_{A(B)} = \frac{1}{2}\left(\idone + \vec{a}_{A(B)}\cdot\hvsigma_{A(B)}\right).
\end{equation}
Here, the vector $ \vec{a}_{A(B)} $ denotes the Bloch vector of the reduced subsystem $ \hat{\varrho}_{A(B)} $, and~$ \hvsigma_{A(B)} $ is the vector of Pauli matrices.

We note that for the two density operators $ \hat{\varrho}_{A} $ and~$ \hat{\varrho}_{B} $, which are derived from the same common pure state $ \ket{\Psi} $, the eigenvalues and thus the lengths of the respective Bloch vectors have to be the same \cite{Nielsen2001}, that is $ \abs{\vec{a}_{A}}=\abs{\vec{a}_{B}} $. These lengths are furthermore related to the concurrence, \cref{eq:def-concurrence0}, by
\begin{equation}
\mathcal{C}=\sqrt{1-\abssq{\vec{a}_{A}}}.
\label{eq:def-concurrence}
\end{equation}

Alternatively, we can relate these lengths to the purity
\begin{align}
\mathcal{P} &\equiv \tr(\hat{\varrho}_{A}^2) 
= \frac{1}{2}\left(1+\abs{\vec{a}_{A}}^2\right),
\label{eq:def-purity}
\end{align}
of the density operator of the subsystem.
From \cref{eq:def-concurrence}, we find the relation
\begin{equation}
\mathcal{P} = 1-\frac{1}{2}\mathcal{C}^2
\label{eq:purity-by-concurrence}
\end{equation}
between the purity and the concurrence.



\subsection{Projective measurements and probabilities}
So far we have concentrated on the state of the combined system. We now analyze measurements on the subsystems.
 
For this purpose we assume that the user makes a projective measurement described by the projection operators
\begin{equation}
\hat{\Pi}_{\vec{e}_{A}}{(a)} \equiv \frac{1}{2}\Big(\idone + (-1)^a\vec{e}_{A}\cdot \hvsigma_{A}\Big)
\label{eq:def-MeasOp_A}
\end{equation}
while the attacker performs a projective measurement given by the operators
\begin{equation}
\hat{\Pi}_{\vec{e}_{B}}{(b)} \equiv \frac{1}{2}\left(\idone + (-1)^b\vec{e}_{B}\cdot \hvsigma_{B}\right),
\label{eq:def-MeasOp_B}
\end{equation}
with $ a=0,1 $ and $ b=0,1 $.

The probability $W_{\vec{e}_{A}}(a)$ to find the bit $ a $ given that the user measures in the direction $\vec{e}_{A}$ and the system is in the state $ \ket{\Psi} $ follows from the Born rule as
\begin{align}
W_{\vec{e}_{A}}(a) = \bra{\Psi}\hat{\Pi}_{\vec{e}_{A}}{(a)}\ket{\Psi}. 
\label{eq:prBb-0}
\end{align}

Analogously, the probability $W_{\vec{e}_{B}}(b)$ to obtain $ b $ provided the attacker measures in the direction $\vec{e}_{B}$ takes the form
\begin{align}
W_{\vec{e}_{B}}(b) = \bra{\Psi}\hat{\Pi}_{\vec{e}_{B}}{(b)}\ket{\Psi} .\label{eq:prAa-0}
\end{align}

By inserting \cref{eq:def-MeasOp_A,eq:def-MeasOp_B} into \cref{eq:prAa-0,eq:prBb-0} respectively, and exploiting \cref{eq:def-rhoaorb,eq:def-rhoaorb-blochv}, we find the marginal probabilities
\begin{equation}\label{eq:prAa-2}
W_{\vec{e}_{A}}(a)= \frac{1}{2}\Big(1+ (-1)^{a}\vec{e}_{A}\cdot\vec{a}_{A}  \Big),
\end{equation}
for the subsystem of the user, and
\begin{equation}\label{eq:prBb-2}
W_{\vec{e}_{B}}(b)= \frac{1}{2}\left(1+ (-1)^{b}\vec{e}_{B}\cdot{\vec{a}_{B}}  \right),
\end{equation}
for the subsystem of the attacker.

The joint probability $W_{\vec{e}_{A},\vec{e}_{B}}(a,b)$ to find the values $a$ and $b$, provided the measurements are in the directions $\vec{e}_{A}$ and $\vec{e}_{B}$, is given by
\begin{align}
W_{\vec{e}_{A},\vec{e}_{B}}(a,b)
\equiv \bra{\Psi} \hat{\Pi}_{\vec{e}_{A}}{(a)}\otimes \hat{\Pi}_{\vec{e}_{B}}{(b)}  \ket{\Psi}
\end{align}
and with the definitions of the projection operators, \cref{eq:def-MeasOp_A,eq:def-MeasOp_B}, this probability takes the form
\begin{align}
W_{\vec{e}_{A},\vec{e}_{B}}(a,b)
= \frac{1}{4}\left(1+(-1)^{a}\vec{e}_{A}\cdot\vec{a}_{A} + (-1)^{b}\vec{e}_{B}\cdot \vec{a}_{B} + (-1)^{a+b}\vec{e}_{A}^{T}\tilde{K}\vec{e}_{B}\right),\label{eq:joint-prob_explicit}
\end{align}
where we have introduced the matrix
\begin{equation}\label{eq:def_K}
\tilde{K} \equiv \bra{\Psi}\hvsigma_{A}\otimes\hvsigma_{B}\ket{\Psi}
\end{equation}
accounting for the correlation between the two subsystems.

\subsection{Bias and correlation}
So far, we have defined the state and the measurement operators for our two-qubit model. We are now in the position to calculate the mutual information for a general pure two qubit state $ \ket{\Psi} $ and projective measurements in both subsystems.

\subsubsection{Definitions}
Inserting the probabilities, \cref{eq:prAa-2,eq:prBb-2,eq:joint-prob_explicit}, back into the definition of the mutual information, \cref{eq:def-mutual_information}, we find
\begin{equation}\label{eq:mI-generic}
I = \frac{1}{4}\sum_{a,b}\left(1 + (-1)^{a}\alpha +(-1)^{b}\beta + (-1)^{a+b}\kappa\right)\log_{2}\left(\frac{1 + (-1)^{a}\alpha +(-1)^{b}\beta + (-1)^{a+b}\kappa}{(1+(-1)^{a}\alpha)(1+(-1)^{b}\beta)}\right)\! ,
\end{equation}
where we have introduced the three parameters 
\begin{equation}\label{eq:def_alphabetakappa}
\alpha\equiv\vec{e}_{A}\cdot\vec{a}_{A} ,\quad \beta\equiv\vec{e}_{B}\cdot\vec{a}_{B} , \quad \kappa\equiv\vec{e}_{A}^{\top}\tilde{K}\vec{e}_{B}.
\end{equation}
Here, $ \alpha $ and $ \beta $ quantify the bias in the measurement outcome on the subsystem~$ A $ and~$ B $, respectively, which can be seen by comparing the definition of these parameters with the marginal probabilities \cref{eq:prAa-2,eq:prBb-2}. Moreover, $ \kappa $ reflects the influence of the correlation between the two subsystems on the joint measurement.

The three parameters are not independent of each other. The bias parameters $ \alpha $ and $ \beta $ both depend on the density operators of their respective subsystem, which are in general not independent, since both derive from a common entangled pure state. The parameter $ \kappa $ also depends on this pure state, as well as on the measurement directions, which also enter in the bias parameters.

In the following we will derive a constraint on these three parameters. For this purpose, we first derive an explicit expression for $ \tilde{K} $

\subsubsection{Constraints}
A general state $ \ket{\Psi} $, given by \cref{eq:state-01basis}, can always be written in the form
\begin{equation}\label{eq:Psi_schmidt}
\ket{\Psi} = \sqrt{ \lambda_{1}}\ket{\uparrow}\ket{\uparrow}+ \sqrt{\lambda_{2}}\ket{\downarrow}\ket{\downarrow},
\end{equation}
due to the Schmidt decomposition \cite{Nielsen2001}, where we have introduced new basis sets $ \lbrace \ket{\uparrow},\ket{\downarrow} \rbrace $ in both subsystems $ A $ and $ B $. Note that in the state $ \ket{\uparrow}\ket{\uparrow} $, in general the spins do not have to point into the same direction anymore.

In \ref{app:Acorrel}, we derive the expression
\begin{equation}\label{eq:K-lambda}
\tilde{K} = \diag(2\sqrt{\lambda_{1}\lambda_{2}},-2\sqrt{\lambda_{1}\lambda_{2}},1).
\end{equation}
for the correlation matrix.

From the definition of the concurrence, \cref{eq:def-concurrence0}, we obtain
\begin{equation}\label{key}
\mathcal{C} = 2\sqrt{\lambda_{1}\lambda_{2}}.
\end{equation}
Together with \cref{eq:def-concurrence} and the normalization condition $ \lambda_{1}+\lambda_{2}=1 $, we arrive at
\begin{equation}\label{eq:lambda1}
\lambda_{1} = \frac{1+\abs{\vec{a}_{A}}}{2}= \frac{1+\sqrt{1-\mathcal{C}^2}}{2} 
\end{equation}
and
\begin{equation}\label{eq:lambda2}
\lambda_{2} = \frac{1-\abs{\vec{a}_{A}}}{2}= \frac{1-\sqrt{1-\mathcal{C}^2}}{2} .
\end{equation}

When we insert \cref{eq:lambda1,eq:lambda2} into the correlation matrix, \cref{eq:def_K}, we obtain
\begin{equation}\label{key}
\tilde{K} = \diag(\mathcal{C},-\mathcal{C},1).
\end{equation}

Furthermore, by calculating the density matrices $ \hat{\varrho}_{A} $ and $ \hat{\varrho}_{B} $ with help of \cref{eq:def-rhoaorb,eq:Psi_schmidt}, and comparing the result with \cref{eq:def-rhoaorb-blochv}, we find $ \vec{a}_{A(B)} = (0,0,\abs{\vec{a}_{A}})^\top $, that is the Bloch vectors point along the $ z $-axis of their respective subsystem.

We are now in the position to calculate the three parameters $ \alpha $, $ \beta $ and $ \kappa $. From their definition, \cref{eq:def_alphabetakappa}, we obtain 
\begin{equation}\label{eq:kappa-explicit}
\kappa = \mathcal{C}\vec{e}_{A,x}\vec{e}_{B,x} - \mathcal{C}\vec{e}_{A,y}\vec{e}_{B,y} + \vec{e}_{A,z}\vec{e}_{B,z}  
\end{equation}
for the correlation parameter, as well as 
\begin{equation}\label{eq:alpha-explicit}
\alpha = \sqrt{1-\mathcal{C}^{2}} \, \vec{e}_{A,z}
\end{equation}
and 
\begin{equation}\label{eq:beta-explicit}
\beta = \sqrt{1-\mathcal{C}^{2}}\, \vec{e}_{B,z}
\end{equation}
for the bias of the user and the attacker, respectively.

In \ref{app:para-const} we prove that \cref{eq:kappa-explicit,eq:alpha-explicit,eq:beta-explicit}  lead to the constraint
\begin{equation}\label{eq:constraint-general}
\frac{1-\mathcal{C}^{2}}{\mathcal{C}^{2}(1-\mathcal{C}^{2}-\alpha^{2})}\kappa^{2} - \frac{2\alpha}{\mathcal{C}^{2}(1-\mathcal{C}^{2}-\alpha^{2})}\kappa\beta + \frac{\mathcal{C}^{2} + \alpha^{2}}{\mathcal{C}^{2}(1-\mathcal{C}^{2}-\alpha^{2})}\beta^{2}  \leq 1.
\end{equation}
For any fixed parameter $ \alpha $, that is for a fixed measurement direction of the user, the equality in \cref{eq:constraint-general} describes an ellipse in the $ \kappa $-$ \beta $-plane. All valid combinations of the parameters $ \beta $ and $ \kappa $ therefore have to lie inside or on the boundary of this ellipse.

\subsubsection{Special cases}
We conclude our discussion by considering the two extreme limits of the concurrence $\mathcal{C}$: (i) a separable bipartite state, and (ii) a maximally entangled state. 
 
For any separable state, that is $ \mathcal{C}=0 $, the constraint becomes
\begin{equation}\label{key}
(\alpha\beta - \kappa)^2 =0,
\end{equation}
which is only fulfilled for $ \kappa = \alpha\beta $. 

As a consequence, we find that the logarithm of \cref{eq:mI-generic} vanishes leading us to
\begin{equation}
I=0,
\label{eq:27}
\end{equation}
as one would expect.

In the other extreme, when the state $ \ket{\Psi} $ is maximally entangled, that is $ \mathcal{C}=1 $, the bias parameters vanish in both subsystems, that is $ \alpha = \beta =0 $, and the correlation is bounded by $ -1\leq \kappa \leq 1 $.

Inserting these values into \cref{eq:mI-generic}, the mutual information takes the form
\begin{equation}\label{key}
I =  \sum_{a,b} \frac{1}{4}\left(1+(-1)^{a+b}\kappa \right)\log_{2}\left(1+(-1)^{a+b}\kappa \right),
\end{equation}
which after performing the summation reads
\begin{align}
I(\kappa)=& \frac{1 + \kappa}{2}   \log_{2}\left(1 + \kappa  \right)+\frac{1 - \kappa }{2}  \log_{2}\left(1 - \kappa  \right).
\label{eq:mI_generic_C=1}
\end{align}
For $\kappa = \pm 1$, we get 
\begin{equation}
 I=1,
\end{equation}
allowing the attacker to obtain complete information about the user's random bit, independent of the user's measurement choice. We emphasize that for a maximally entangled state the user cannot prevent the attacker from finding out his random bit. 
%

\section{Worst-case scenario}
\label{sec:wcs}
In the preceding section we have derived a general expression for the mutual information of a two-qubit system which depends on the concurrence and the measurements performed relatively to the reduced density matrices on both subsystems.
We now discuss special measurement strategies of user and attacker and highlight the important role of entanglement in our scheme. Throughout this section we consider the worst case for the user, that is the attacker somehow knows the user's measurement directions, as well as the complete state $\ket{\Psi}$.

\subsection{User's choice of measurement direction}
For a QRNG, a user would naturally maximize the entropy of the bits and therefore choose his measurements in such a way that he obtains uniformly distributed bits with 
\begin{equation}
W_{\vec{e}_{A}}{(0)}=W_{\vec{e}_{A}}{(1)}=\frac{1}{2}.
\label{eq:prob-equal-dist}
\end{equation}

According to \cref{eq:prAa-2} this requirement translates into condition
\begin{equation}
\alpha=\vec{e}_{A}\cdot \vec{a}_{A}=0
\label{eq:meas-direction_equal-dist}
\end{equation}
for the user's measurement. 

Geometrically, this prescription means $ \vec{e}_{A}\perp \vec{a}_{A} $, that is the measurement is perpendicular to the Bloch vector of $\hat{\varrho}_{A}$. There are infinitely many vectors $ \vec{e}_{A} $ that fulfill this condition. Throughout this section, we consider this situation with a fixed $ \vec{e}_{A} $ but generalize it slightly in \ref{sec:rm} by allowing random measurements corresponding to two different  $\vec{e}_{A}$, which are both perpendicular to $ \vec{a}_{A} $. 

When we substitute \cref{eq:meas-direction_equal-dist} into \cref{eq:mI-generic}, we obtain the mutual information
\begin{equation}
I = \frac{1}{4}\sum_{a,b} \left(1+ (-1)^{b}\beta +(-1)^{a+b}\kappa\right) \log_{2}\left(1+(-1)^{a+b}\frac{\kappa}{1+(-1)^{b}\beta}\right).
\label{eq:mI_thetaA=pi/2}
\end{equation}
The parameters $ \kappa $ and $ \beta $ are not independent, but constrained by the equation
\begin{equation}\label{eq:constraint-k-s}
\left(\frac{\kappa}{\mathcal{C}}\right)^{2} + \left(\frac{\beta}{\sqrt{1-\mathcal{C}^{2}}}\right)^{2} \leq 1
\end{equation}
corresponding to an ellipse with the semi-major and semi-minor axes coinciding with the $ \kappa $ and $ \beta $ axes, which follows directly from \cref{eq:constraint-general} for $ \alpha =0 $. 


\subsection{Maximum of mutual information}
In order to guarantee the secrecy of his random bits, the user has to address the question: What is the maximal information following from \cref{eq:mI_thetaA=pi/2} any attacker can obtain about the bit $a$ for the given setting?

\subsubsection{Exact expression}
Since the mutual information is a convex function in the $ \kappa $-$ \beta $-plane, its maximum has to lie on the boundary of the ellipse. 
\begin{figure}
\centering
	\begin{minipage}[t]{.45\textwidth}
		a)\\ \includegraphics[scale=0.14]{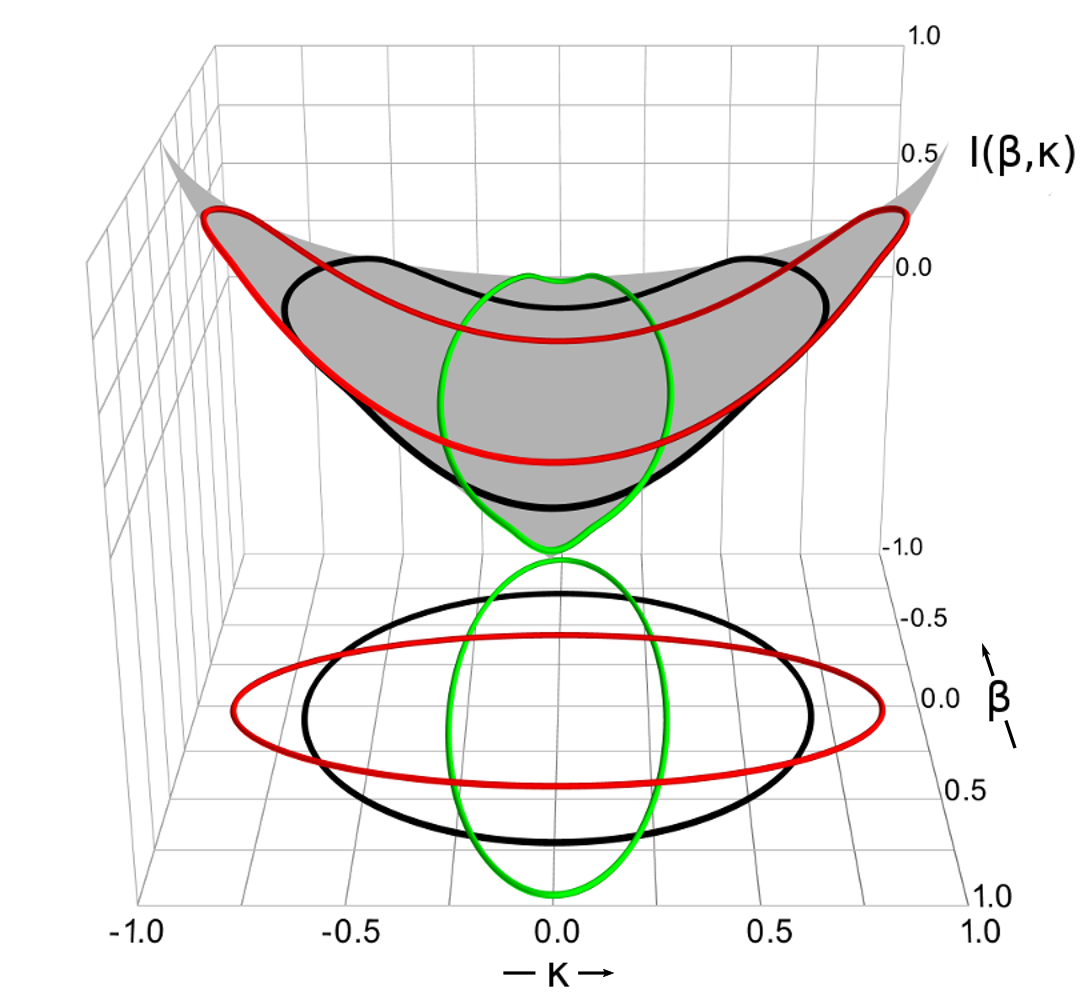}
	\end{minipage}
	\begin{minipage}[t]{.50\textwidth}
		\hspace{-3cm}b)\\ 
		\centering
		 \includegraphics[scale=0.20]{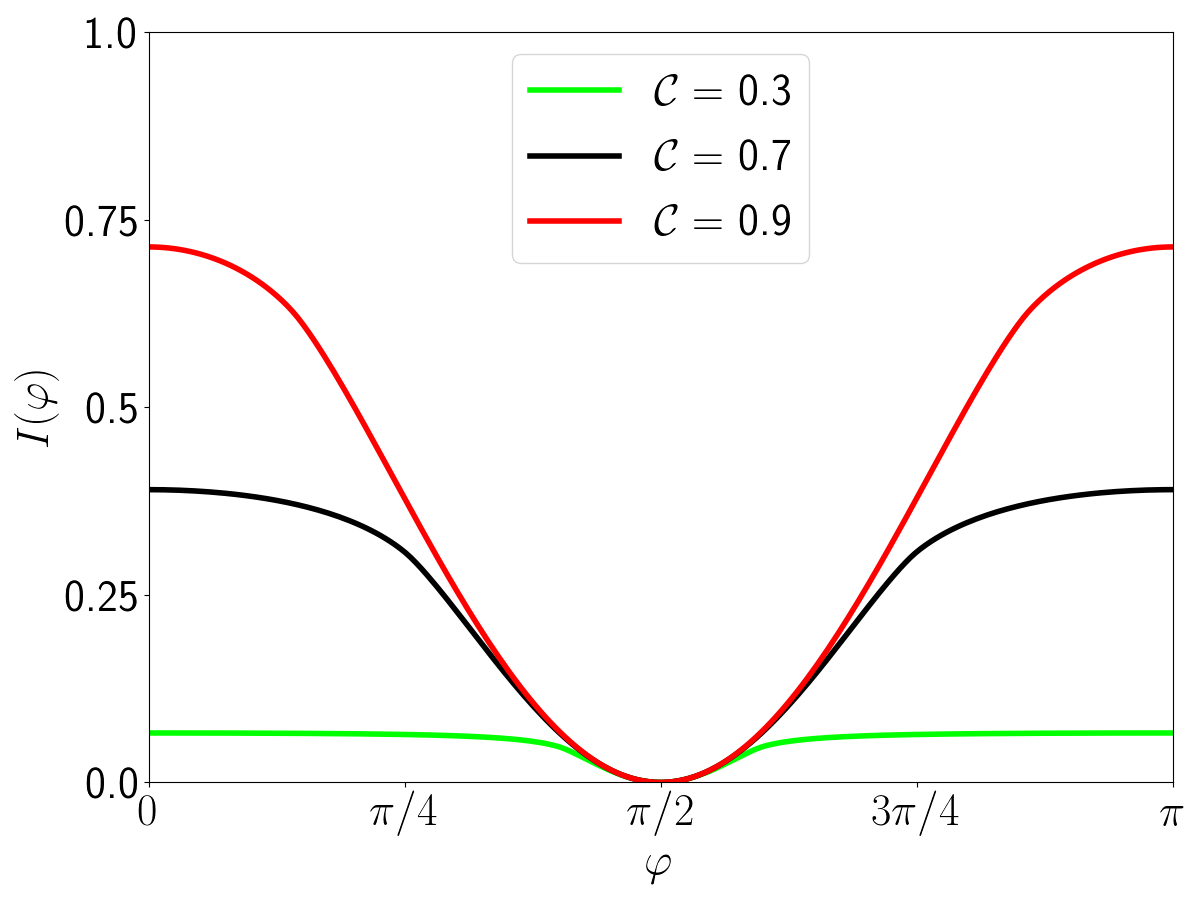}
	\end{minipage}
\caption{Geometric determination of the absolute maximum of the mutual information $I$ according to \cref{eq:mI_thetaA=pi/2} under the constraint \cref{eq:constraint-k-s}. a) The mutual information (top) is shown in its dependence on the correlation $ \kappa $ of the two systems and the bias $ \beta $ in the measurement of the attacker.  
The ellipses in the $ \kappa $-$ \beta $-plane (bottom) enclose all the possible combinations of $ \kappa $ and $ \beta $ that can be achieved by any measurement direction $\vec{e}_{B}$ of the attacker. The eccentricities of these ellipses are determined solely by the concurrence $\mathcal{C}$ quantifying the degree of entanglement between the qubits of the user and the attacker. The green, black and red ellipses correspond to $ \mathcal{C}=0.3, \mathcal{C}=0.7 $ and $ \mathcal{C}=0.9 $, respectively.
Due to the shape of the mutual information, its maximal value is found on the intersection between the ellipse and the $ \kappa $-axis, independent of the concurrence.
For increasing concurrences $ \mathcal{C} $ the mutual information at this intersection increases. Thus, the maximal mutual information increases with increasing concurrence.
b) Mutual information along the ellipses parameterized by an angle $ \varphi $ and corresponding to the same values of the concurrences $ \mathcal{C} $ as in a). The angle $ \varphi $ is chosen such that $ \varphi=0 $ corresponds to the intersection between the ellipse and the positive $ \kappa $-axis. For symmetry reasons, we only parameterize the ellipse from $ \varphi=0 $ to $ \varphi=\pi $. 
The mutual information is maximal for the attacker choosing his measurement for the parameter $ \varphi = 0 $ or $ \varphi = \pi $, that is at the intersections of the ellipse with the $ \kappa $-axis, independent of the concurrence $ \mathcal{C} $. }
\label{fig:mu-inf-contour-onemeaseach}
\end{figure}
In \cref{fig:mu-inf-contour-onemeaseach} we show that the mutual information is maximized on the intersection of the ellipse given by the constraint, \cref{eq:constraint-k-s}, and the $ \kappa $-axis. These points lead to the two conditions  
\begin{equation}\label{eq:beta-maxI}
\beta =0
\end{equation}
and
\begin{equation}\label{eq:kappa-maxI}
\kappa = \pm \mathcal{C}.
\end{equation}

The condition on the attacker's bias, \cref{eq:beta-maxI}, means that the measurement direction of the attacker $ \vec{e}_{B} $ is perpendicular to the Bloch vector $ \vec{a}_{B} $ of his subsystem. Hence, the attacker will also obtain a uniform distribution of his bits. As for the user, there are infinitely many measurement directions, which fulfill this condition.

The second condition, \cref{eq:kappa-maxI}, together with \cref{eq:kappa-explicit,eq:alpha-explicit,eq:meas-direction_equal-dist}, poses the requirement
\begin{equation}\label{key}
\vec{e}_{A,x}\vec{e}_{B,x} - \vec{e}_{A,y}\vec{e}_{B,y} = \pm 1
\end{equation} 
on the choice of the attacker's measurement, which restricts the attacker's measurement to two directions. He can either choose $ \vec{e}_{B}= (\vec{e}_{A,x},-\vec{e}_{A,y},0) $ or $ \vec{e}_{B}= (-\vec{e}_{A,x},\vec{e}_{A,y},0) $. 

As a result, by inserting \cref{eq:beta-maxI,eq:kappa-maxI} into \cref{eq:mI_thetaA=pi/2}, we find
\begin{equation}\label{key}
I_{\max} = \frac{1}{4}\sum_{a,b} \left(1+(-1)^{a+b} \mathcal{C}\right) \log_{2}\left(1+(-1)^{a+b}\mathcal{C} \right),
\end{equation}
and after performing the summations the maximal mutual information an attacker can gain by performing a measurement on the environment reads
\begin{align}
I_{\max}
= \frac{ 1 + \mathcal{C}}{2} \log_{2}\left(1 + \mathcal{C} \right)  +  \frac{1 - \mathcal{C}}{2} \log_{2}\left(1 -\mathcal{C} \right).
\label{eq:Imax_for_chi=pihalf}
\end{align}
This expression is the central result of our article. We note, that we can also find the this result analytically. This rather lengthy calculation is shown in detail in \ref{app:maxI}.

It is interesting to note that a similar equation holds true if the user switches between different measurements. In \ref{sec:rm} we discuss this scenario in detail.

\begin{figure}
\centering
\includegraphics[scale=0.25]{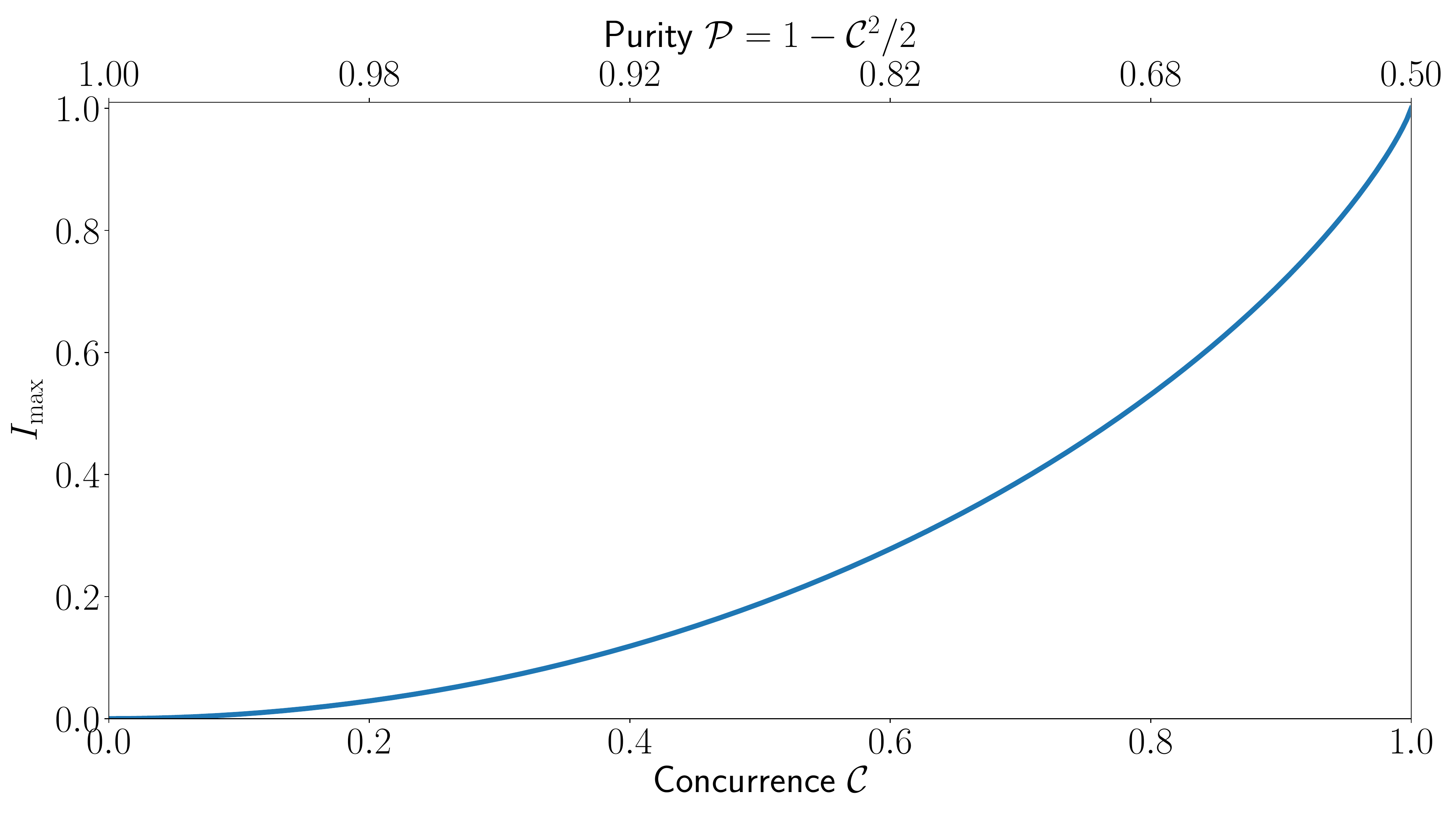}
\caption{Worst-case scenario for the user who chooses a measurement such that he obtains uniformly distributed bits. The attacker selects his measurements as to maximize the mutual information. The corresponding mutual information $I_{\max}$ increases for increasing values of the concurrence $\mathcal{C}$ (horizontal axis on the bottom) and decreases with increasing purity $\mathcal{P}$ of the state $\hat{\varrho}_{A}$ of the user (horizontal axis on the top). Close to a pure state, that is $\mathcal{P}=1$, the decrease is linear.}
\label{fig:maximum-distribution}
\end{figure} 
Figure \ref{fig:maximum-distribution} shows the maximal mutual information, \cref{eq:Imax_for_chi=pihalf}, in its dependence on both the concurrence and the purity.
The more the two systems are entangled, that is the less pure the state of the user, the more information can be gained from one measurement result about the other.

\subsubsection{Asymptotic expressions}
If the complete state $\ket{\Psi}$ is only weakly entangled corresponding to $C \ll 1$, we can perform a Taylor expansion  
\begin{equation}
\ln(1\pm x) \cong \pm x + x^2/2 + \mathcal{O}(x^3),
\end{equation} 
of the logarithm to second order and thus approximate \cref{eq:Imax_for_chi=pihalf} by
\begin{align}
I_{\max}&\cong 
\frac{\mathcal{C}^{2}}{2\ln2}+\mathcal{O}(\mathcal{C}^3). \label{eq:max-taylor-smallC}
\end{align}
Hence, for small concurrences $\mathcal{C}$ the maximal mutual information only grows \textit{quadratically}, and there is almost no mutual information. The additional information on the more probable bit is almost compensated by the less information about the less probable bit. Thus, for small concurrences $\mathcal{C}$, the information an attacker can gain is almost negligible, providing a certain robustness of such a QRNG scheme against small entanglement between the QRNG's system and the environment. 

From the viewpoint of the user, \cref{eq:max-taylor-smallC} means that the mutual information decreases linearly with the purity for $\mathcal{P}\lesssim 1$. Indeed, when we substitute the connection, \cref{eq:purity-by-concurrence} between $ \mathcal{P} $ and $ \mathcal{C}^2 $ into \cref{eq:max-taylor-smallC} we find
\begin{equation}\label{key}
I_{\max} \cong \frac{1}{\ln 2}\left(1 - \mathcal{P} \right).
\end{equation}

On the other hand, for values of $\mathcal{C} \lesssim 1$ the mutual information grows rapidly with increasing $\mathcal{C}$, since the positive term in \cref{eq:Imax_for_chi=pihalf} is weighted with a high probability, while the factor decreasing the mutual information is far less probable.

We finally remark that in our scheme the user needs to know the state $\hat{\varrho}_{A}$ of his subsystem, which in general can be obtained by state tomography. The connection, \cref{eq:purity-by-concurrence}, between the concurrence and the purity of the user's subsystem then allows the user to find an upper bound on the privacy of his data.  
 

\subsection{Binary entropy}
We note that \cref{eq:Imax_for_chi=pihalf} enjoys an elementary interpretation, 
%
based on the binary entropy
\begin{equation}
H_{b}(p) \equiv -p\log_{2}\left(p\right)  - (1 - p)\log_{2}\left(1-p\right),
\end{equation}
for a probability $p$. Indeed, \cref{eq:Imax_for_chi=pihalf} can be written as
\begin{equation}
I_{\max}= 1 - H_{b}\left(\frac{1+\mathcal{C}}{2}\right).
\end{equation}
The first term on the right-hand side corresponds to the entropy of the user's random number without any correlation to another measurement result. This value is one, due to the fact that the user's bit is equally distributed. 

The second term on the right-hand side, which subtracts from the user's entropy, is the conditional entropy of the user's bit, when the attacker's bit is known. This contribution corresponds to the entropy that remains, even when the attacker has made a measurement, and therefore reduces the information he can gain. Interestingly, this entropy corresponds to a binary entropy, with probabilities \begin{equation}\label{key}
p_{\pm}\equiv\frac{1}{2}\left(1\pm\mathcal{C}\right).
\end{equation} 

Hence, the concurrence $\mathcal{C}$ is a measure of the deviation from a uniform binary distribution. For a vanishing concurrence the user's bit is equally likely for any value of the attacker's bit, while with increasing concurrence the probability of having coincidental results between the user's and the attacker's outcome increases.

\subsection{Privacy of the quantum random numbers and quantum state discrimination}
We conclude our discussion of the worst case scenario by taking a different point of view on the privacy of the random numbers generated by a QRNG. Indeed the question of how much information an attacker can maximally gain can also be considered as a quantum state discrimination task \cite{Helstrom1976,Bergou2004,Bergou2010}. By performing a measurement on the subsystem $A$, the state of the attacker in the subsystem $ B $ is a pure state, depending on the outcome $ a $ of the measurement performed on the subsystem $ A $. The task of the attacker is to discriminate his two states.

When the two states are orthogonal, the attacker can always perform a measurement, which allows him to discriminate between the two states with certainty. In general, however, the two states are not orthogonal and therefore there is no measurement that can decide unambiguously between the two cases. 

It is well known, that the maximal mutual information accessible in this case is bounded from above and below by the inequalities
\begin{equation}\label{eq:imax-bounds}
\chi_{\mathrm{JRW}}\leq I_{\max}\leq \chi_H. 
\end{equation}

The upper bound is the well known Holevo bound \cite{Nielsen2001}
\begin{equation}\label{eq:eq-Holevo-def}
\chi_H \equiv S(\hat{\varrho}_{B}) - \sum_{a}W_{\vec{e}_{A}}(a)S(\hat{\varrho}_{B|a})
\end{equation}
with $ \hat{\varrho}_{B|a}\equiv \ket{\psi_{a}}_{B}\bra{\psi_{a}} $ and the Shannon entropy
\begin{equation}\label{eq:shannon-entropy}
S(\hat{\varrho})= -\tr(\hat{\varrho}\log_{2}\hat{\varrho})= -\sum_{k}\lambda_{k}\log_{2}\lambda_{k},
\end{equation}
where $ \lambda_{k} $ denote the eigenvalues of the density operator $ \hat{\varrho} $.

The lower bound for the maximal accessible information, proposed by Josza, Robb and Wootters \cite{Jozsa1994}, is given by 
\begin{equation}\label{eq:JRW-def}
\chi_{\mathrm{JRW}}\equiv Q(\hat{\varrho}_{B}) - \sum_{a}W_{\vec{e}_{A}}(a)Q(\hat{\varrho}_{B|a})
\end{equation}
with the subentropy
\begin{equation}\label{eq:subentropy}
Q(\hat{\varrho}) \equiv -\sum_{k}\left(\sum_{l\neq k} \frac{\lambda_{k}}{\lambda_{k} - \lambda_{l}}\right)\lambda_{k}\log_{2}\lambda_{k}.
\end{equation}

We now consider the state discrimination task for our problem of the QRNG in the worst-case scenario. As a first step, we show that the states the attacker obtains are not orthogonal, as long as the combined state $ \ket{\Psi} $, defined in \cref{eq:state-01basis}, is not maximally entangled.

For the measurement outcome $ a $, the user finds the state 
\begin{equation}\label{key}
\ket{\psi_{a}}_{A}= \frac{1}{\sqrt{2}}\left(\ket{\uparrow}_{A} +(-1)^{a}  \e^{\ii\varphi} \ket{\downarrow}_{A}\right),
\end{equation}
with an arbitrary but fixed phase $ \varphi $.

Therefore the state $ \ket{\psi_{a}}_{B} $ in the subsystem $ B $, conditioned on the measurement result $ a $, reads
\begin{equation}\label{key}
\ket{\psi_{a}}_{B} =\frac{ {}_{A}\braket{\psi_{a}}{\Psi}}{\sqrt{W_{\vec{e}_{A}}(a)}} = \sqrt{2}\, {}_{A}\braket{\psi_{a}}{\Psi},
\end{equation}
where the probability $ W_{\vec{e}_{A}}(a) = 1/2 $, given by \cref{eq:prAa-0}, in the denominator ensures normalization.

We recall the state $ \ket{\Psi} $ in the Schmidt decomposition, \cref{eq:Psi_schmidt}, and find
\begin{equation}\label{eq:psiaB}
\ket{\psi_{a}}_{B} = \sqrt{\frac{1+\abs{a_{A}}}{2}}\ket{\uparrow}_{B} +(-1)^{a} \sqrt{\frac{1-\abs{a_{A}}}{2}}\e^{-\ii\varphi}\ket{\downarrow}_{B}
\end{equation}
for the state in the subsystem $ B $, conditioned that the user has measured the bit $ a $. 

For $ \abs{\vec{a}_{A}} >0$ the scalar product 
\begin{equation}\label{key}
{}_{B}\braket{\psi_{0}}{\psi_{1}}_{B}= \frac{1+\abs{\vec{a}_{A}}}{2} - \frac{1-\abs{\vec{a}_{A}}}{2}= \abs{\vec{a}_{A}}
\end{equation}
between the two states $ \ket{\psi_{0}}_{B} $ and $ \ket{\psi_{1}}_{B} $, following from \cref{eq:psiaB}, does not vanish, and these two states are not orthogonal. 

In the next step, we calculate the bounds given by \cref{eq:eq-Holevo-def,eq:JRW-def}. Since the entropy vanishes for a pure state, the Holevo bound is given by the Shannon entropy of the state
$\hat{\varrho}_{B} $ of the attacker $ S(\hat{\varrho}_{B}) $. 

With the explicit formulas  \cref{eq:lambda1,eq:lambda2} for the eigenvalues $ \lambda_{k} $ and the definition of the Shannon entropy $ S(\hat{\varrho}) $, \cref{eq:shannon-entropy}, we find
\begin{equation}\label{eq:holevo-result}
\chi_{H} = -\frac{1+\sqrt{1-\mathcal{C}^2}}{2}\log_{2}\left(\frac{1+\sqrt{1-\mathcal{C}^2}}{2}\right)-\frac{1-\sqrt{1-\mathcal{C}^2}}{2}\log_{2}\left(\frac{1-\sqrt{1-\mathcal{C}^2}}{2}\right)
\end{equation}
for the Holevo bound.

Since the subentropy also vanishes for pure states, the maximal accessible information is given by the subentropy $ Q(\hat{\varrho}_{B}) $ of the attacker's density matrix. By using the eigenvalues, \cref{eq:lambda1,eq:lambda2}, of this state, together with the definition of $ Q(\hat{\varrho}) $, \cref{eq:subentropy}, we obtain
\begin{equation}\label{eq:min-access-info}
\chi_{\mathrm{JRW}} =- \frac{(1+\sqrt{1-\mathcal{C}^2})^2}{4\sqrt{1-\mathcal{C}^2}}\log_{2}\left(\frac{1+\sqrt{1-\mathcal{C}^2}}{2}\right) +\frac{(1-\sqrt{1-\mathcal{C}^2})^2}{4\sqrt{1-\mathcal{C}^2}}\log_{2}\left(\frac{1-\sqrt{1-\mathcal{C}^2}}{2}\right)
\end{equation}
for the maximal accessible information.

\begin{figure}
	\centering
	\includegraphics[scale=.25]{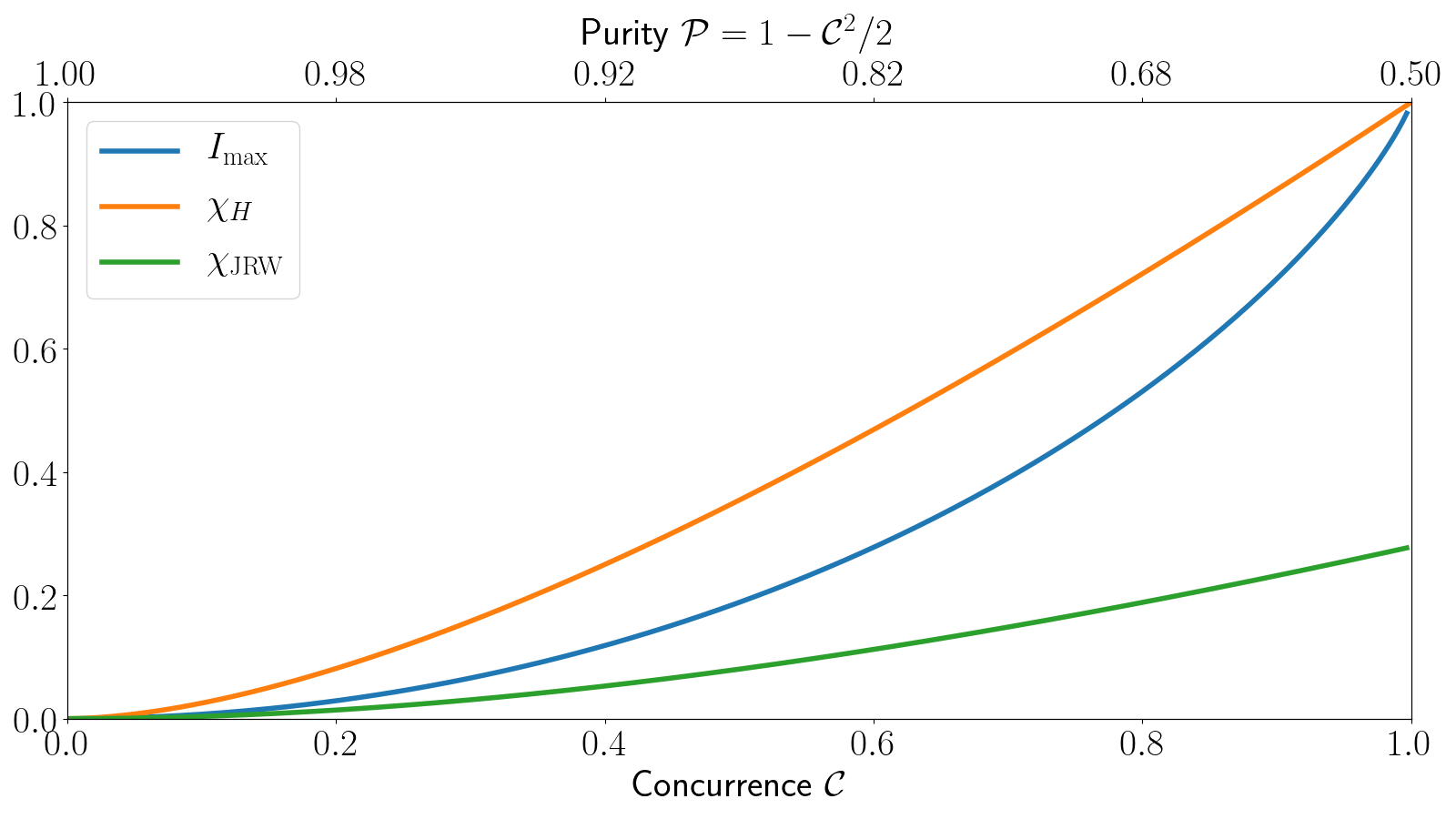}
\caption{Comparison between the maximal mutual information $ I_{\max} $, \cref{eq:Imax_for_chi=pihalf}, the Holevo bound $ \chi_{H} $, \cref{eq:holevo-result}, and the lower bound $ \chi_{\mathrm{JRW}} $ for the maximal mutual information accessible, \cref{eq:min-access-info}. The maximal mutual information for a projective measurement lies between the Holevo bound and the lower bound for the maximal mutual information for all values of the concurrence except as one would expect. Except for the boundaries $ \mathcal{C}=0 $ and $ \mathcal{C}=1 $ the mutual information is strictly lower than the Holevo bound. }
\label{fig:bounds}
\end{figure}
In \cref{fig:bounds} we compare our result for the maximal mutual information, \cref{eq:Imax_for_chi=pihalf}, with the Holevo bound, \cref{eq:holevo-result}, and the minimal accessible information, \cref{eq:min-access-info}. 
The result of our worst case considerations, \cref{eq:Imax_for_chi=pihalf}, is thus between the two bounds as expected. However, our result is strictly lower than the Holevo bound except for the boundary values $ \mathcal{C}=0 $ and $ \mathcal{C}=1 $, and therefore an improvement for the user over just assuming the Holevo bound. 
This advantage originates from the fact, that the Holevo bound is only dependent on the maximal information contained of the state $ \hat{\varrho}_{B} $ in the subsystem~$ B $, independent of the composition of this state, that is of exact form of the states~$ \ket{\psi_{0}}_{B} $ and~$ \ket{\psi_{1}}_{B} $. The Holevo bound is only tight if $ \ket{\psi_{0}}_{B} $ and $ \ket{\psi_{1}}_{B} $ are identical or orthogonal, which is only fulfilled if the pure state $ \ket{\Psi} $ of the combined system is either separable or maximally entangled. In all the cases in between the Holevo bound is cannot tight. Our result, \cref{eq:Imax_for_chi=pihalf}, is exact, and therefore takes the measurement of the user and hence the exact form of $ \ket{\psi_{0}}_{B} $ and $ \ket{\psi_{1}}_{B} $ into account.

\section{Conclusions and Outlook} \label{sec:conclusionsandoutlook}
We are now in the position to summarize our results and provide a short outlook.
Throughout this article we have discussed the privacy of random numbers created by a  non-ideal QRNG represented by a single qubit system coupled to another qubit system that models the environment an attacker may have access to and which is due to the fact that the user cannot prepare a perfectly pure quantum state. 
%
   
We have provided an upper bound, \cref{eq:Imax_for_chi=pihalf}, on how much information the attacker can gain about the user's random bit. From this expression, we conclude that the limiting factor on this bound is the entanglement between the QRNG system and its environment, quantified by the concurrence.   
We emphasize that our upper bound holds without any further restrictions on the user's or attacker's measurement scheme.


Moreover, we have shown that our scheme can be interpreted in terms of quantum state discrimination. This point of view allows us to compare the result to the known bounds. Since our worst case analysis is exact, our result improves the well-known Holevo bound in this special case.


We emphasize that our results can directly be applied to different QRNG realizations. Furthermore, our analysis can be extended to generalized measurements, such as POVMs, and measurement strategies, which may lead to a further reduction of the maximal mutual information. This extension also allows us to include the effects of detector efficiencies into our existing model. 

With these modifications our model will constitute an elementary yet useful tool to estimate the maximal information the attacker can gain on the numbers created by QRNGs. We will also be able to extend our model to self-testing QRNG devices, by further including the state tomography directly into the measurement protocol. Finally we might improve existing lower bounds on the min-entropy. These topics, however, go beyond the scope of the present article and will be addressed in a future publication.

\section*{Acknowledgments}
We are grateful to A. Friedrich, E. Giese, M. Steiner, A. Wolf and S. Wölk for many fruitful discussions. We thank M. Beck for sending us Ref.~\cite{Cutshall2020} before publication.
J.S. thanks the Center for Integrated Quantum Science  and  Technology  (IQ\textsuperscript{ST})  for  a  fellowship within the framework of the Quantum Alliance sponsored by the Ministry of Science, Research and Arts, Baden-W{\"u}rttemberg. T.S. acknowledges support from the EU Quantum Flagship project QRANGE (grant no. 820405). W.P.S. is grateful to Texas A{\&}M University for a Faculty Fellowship at the Hagler Institute for Advanced Study at Texas A{\&}M University and to Texas A{\&}M AgriLife Research for the support of this work. The research of IQ\textsuperscript{ST} is financially supported by the Ministry of Science, Research and Arts, Baden-W{\"u}rttemberg.

\appendix

\section{Calculation of the correlation matrix}\label{app:Acorrel}
In this Appendix we will calculate the correlation matrix $ K $, defined in \cref{eq:def_K}, for a general entangled two qubit state $ \ket{\Psi} $. 

We start from the state
\begin{equation}\label{eq:appA-state-schmidt}
\ket{\Psi} = \sqrt{ \lambda_{1}}\ket{\uparrow\uparrow}+ \sqrt{\lambda_{2}}\ket{\downarrow\downarrow},
\end{equation}
as defined in \cref{eq:Psi_schmidt}. Since this state is symmetric in the two subsystems, it is obvious that the matrix $ \tilde{K} $ has to be symmetric too, that is $ \tilde{K}_{ij} = \tilde{K}_{ji} $.

Thus, we only have to evaluate six coefficients. We start with the three off-diagonal coefficients. 
The first one is
\begin{equation}\label{key}
\tilde{K}_{xy} = \bra{\Psi}\hat{\sigma}_{x}\otimes\hat{\sigma}_{y}\ket{\Psi}.
\end{equation}
By inserting the definition of the state, \cref{eq:appA-state-schmidt}, as well as of the Pauli matrices, we obtain
\begin{equation}\label{key}
\tilde{K}_{xy} = \bra{\Psi}\left(\ii \sqrt{\lambda_{1}}\ket{\uparrow\uparrow} - \ii\sqrt{\lambda_{2}} \ket{\downarrow\downarrow}\right),
\end{equation}
which then becomes
\begin{equation}\label{key}
\tilde{K}_{xy} = \ii \left( \sqrt{\lambda_{1}\lambda_{2}} - \ii\sqrt{\lambda_{2}\lambda_{1}} \right)=0.
\end{equation}
Furthermore, in the case of $ i=x,y $ and $ j=z $, we find
\begin{equation}\label{key}
\hat{\sigma}_{i} \otimes \hat{\sigma}_{z} \ket{\Psi} = c_{i,1}\ket{\uparrow\downarrow} + c_{i,2}\ket{\downarrow\uparrow}
\end{equation}
with some coefficients $ c_{i,1} $ and $ c_{i,2} $, depending on $ i=x,y $. These states are clearly orthogonal to the state $ \ket{\Psi} $, and therefore we find $ \tilde{K}_{xz}=\tilde{K}_{yz}=0 $.

Hence, the correlation matrix is diagonal in the Schmidt basis. The only remaining task is therefore to find the diagonal components.
For $ i=j=x $ we find 
\begin{equation}\label{key}
\tilde{K}_{xx} = \bra{\Psi}\left( \sqrt{\lambda_{2}}\ket{\uparrow\uparrow} + \sqrt{\lambda_{1}} \ket{\downarrow\downarrow}\right)
\end{equation} 
which gives
\begin{equation}\label{key}
\tilde{K}_{xx} = 2\sqrt{\lambda_{1}\lambda_{2}}.
\end{equation}

Analogously, for $ i=j=y $, we have
\begin{equation}\label{key}
\tilde{K}_{xx} = \bra{\Psi}\left( -\sqrt{\lambda_{2}}\ket{\uparrow\uparrow} - \sqrt{\lambda_{1}} \ket{\downarrow\downarrow}\right)
\end{equation} 
leading to
\begin{equation}\label{key}
\tilde{K}_{xx} = -2\sqrt{\lambda_{1}\lambda_{2}}.
\end{equation}

Finally, for the case $ i=j=z $ we find
\begin{equation}\label{key}
\tilde{K}_{xx} = \bra{\Psi}\left( \sqrt{\lambda_{1}}\ket{\uparrow\uparrow} + \sqrt{\lambda_{2}} \ket{\downarrow\downarrow}\right)= \braket{\Psi}{\Psi}=1,
\end{equation} 
since the state $ \ket{\Psi} $ is normalized.

Combining all of the above results, we finally obtain the correlation matrix 
\begin{equation}\label{eq:appA_K-lambda}
\tilde{K} = \diag(2\sqrt{\lambda_{1}\lambda_{2}},-2\sqrt{\lambda_{1}\lambda_{2}},1).
\end{equation}

\section{Parameter constraints}\label{app:para-const}
In this Appendix, we derive the constraints for the parameters $ \alpha, \beta $ and $ \gamma $ for a general state $ \ket{\Psi} $. In fact, we show that for an arbitrary but fixed measurement parameter $ \alpha $ the two parameters $ \beta $ and $ \kappa $ lie inside an ellipse in the $ \kappa $-$ \beta $-plane, while the shape of the ellipse is determined by $ \alpha $.

We have shown in the main article that the three parameters are given by
\begin{equation}\label{eq:appA-kappa-explicit}
\kappa = \mathcal{C}\vec{e}_{A,x}\vec{e}_{B,x} - \mathcal{C}\vec{e}_{A,y}\vec{e}_{B,y} + \vec{e}_{A,z}\vec{e}_{B,z}  
\end{equation}
as well as 
\begin{equation}\label{eq:appA-alpha-explicit}
\alpha = \sqrt{1-\mathcal{C}^{2}}\vec{e}_{A,z}
\end{equation}
and 
\begin{equation}\label{eq:appA-beta-explicit}
\beta = \sqrt{1-\mathcal{C}^{2}}\vec{e}_{B,z}.
\end{equation}

By introducing spherical coordinates in both subsystems $ A $ and $ B $, that is 
\begin{equation}\label{key}
\vec{e}_{A(B)}= \begin{pmatrix}
\sin\theta_{A(B)}\cos\varphi_{A(B)}\\
\sin\theta_{A(B)}\sin\varphi_{A(B)}\\
\cos\theta_{A(B)}
\end{pmatrix},
\end{equation}
the parameters of \cref{eq:appA-alpha-explicit,eq:appA-beta-explicit,eq:appA-kappa-explicit} can be rewritten as
\begin{equation}\label{eq:appA-kappa-sc}
\kappa = \mathcal{C}\sin\theta_{A}\sin\theta_{B}\cos(\varphi_{A}-\varphi_{B})+\cos\theta_{A}\cos\theta_{B}
\end{equation}
as well as 
\begin{equation}\label{eq:appA-alpha-sc}
\alpha = \sqrt{1-\mathcal{C}^{2}}\cos\theta_{A}
\end{equation}
and
\begin{equation}\label{eq:appA-beta-sc}
\beta = \sqrt{1-\mathcal{C}^{2}}\cos\theta_{B}.
\end{equation}

From \cref{eq:appA-kappa-sc} we get
\begin{equation}\label{eq:appA-kappa-sc2}
\left(\kappa - \cos\theta_{A}\cos\theta_{B}\right)^{2} = \mathcal{C}^{2}\sin^{2}\theta_{A}\sin^{2}\theta_{B}\cos^{2}(\varphi_{A}-\varphi_{B})
\end{equation}
by bringing the second term on the right hand side of \cref{eq:appA-kappa-sc} to the left hand side 
and squaring the resulting equation. Since we have $ \cos x \leq 1 $ for all $ x $, we furthermore find
\begin{equation}\label{key}
\left(\kappa - \cos\theta_{A}\cos\theta_{B}\right)^{2} \leq \mathcal{C}^{2}\sin^{2}\theta_{A}\sin^{2}\theta_{B},
\end{equation}
which is equivalent to
\begin{equation}\label{eq:appB-kappa2}
\kappa^2 - 2\cos\theta_{A}\cos\theta_{B}\kappa + \cos^2\theta_{A}\cos^2\theta_{B} \leq \mathcal{C}^2(1-\cos^2\theta_{A})( 1-\cos^2\theta_{B}).
\end{equation}
Solving \cref{eq:appA-alpha-sc,eq:appA-beta-sc} for $ \cos\theta_{A} $ and $ \cos\theta_{B} $, respectively, and inserting these relations into \cref{eq:appB-kappa2} gives
\begin{equation}\label{key}
\kappa^2 - \frac{2}{1-\mathcal{C}^2}{\alpha}{\beta}\kappa + \frac{{\alpha}^2{\beta}^2}{1-\mathcal{C}^2} \leq \frac{\mathcal{C}^2 }{(1-\mathcal{C}^2)^2}\left(1 - \mathcal{C}^2-{\alpha}^2\right)\left(1 - \mathcal{C}^2-{\beta}^2\right),
\end{equation}
which can be rewritten as
\begin{equation}\label{key}
\frac{1-\mathcal{C}^2}{\mathcal{C}^2(1 - \mathcal{C}^2-{\alpha}^2)}\kappa^2 - \frac{2{\alpha}}{\mathcal{C}^2(1 - \mathcal{C}^2-{\alpha}^2)}{\beta}\kappa + \frac{{\alpha}^2 + \mathcal{C}^2}{\mathcal{C}^2(1 - \mathcal{C}^2-{\alpha}^2)}{\beta}^2 \leq 1.
\end{equation}
Note, that for a fixed parameter $ \alpha $, this inequality describes the area enclosed by an ellipse in the $ \kappa $-$ \beta $-plane, where the shape and orientation of the ellipse are determined by $ \alpha $ and the concurrence $ \mathcal{C} $.

\section{Maximizing the mutual information}\label{app:maxI}
In this Appendix, we analytically derive the maximal mutual information an attacker can have access to, in the case of a QRNG setting. The measurement of the user is described by a vector $ \vec{e}_{A} $ with $ \vec{e}_{A}\cdot\vec{a}_{A} =0$. 

The mutual information for this setting is given by
\begin{equation}\label{eq:appB_I-ks}
I = \frac{1}{4}\sum_{a,b}\left(1+(-1)^{b}\beta + (-1)^{a+b}\kappa\right)\log_{2}\left(\frac{1+(-1)^{b}\beta + (-1)^{a+b}\kappa}{1+(-1)^{b}\beta}\right),
\end{equation}
while the two parameters $ \kappa $ and $ \beta $ are constraint by 
\begin{equation}\label{key}
\left(\frac{\kappa}{\mathcal{C}}\right)^{2} + \left(\frac{\beta}{\sqrt{1-\mathcal{C}^{2}}}\right)^{2} \leq 1,
\end{equation}
which means that they lie inside an ellipse in the $ \kappa $-$ \beta $-plane.

\subsection{Convexity}
It is well known that the mutual information is convex as a function of the conditional probability $ W_{\vec{e}_{A},\vec{e}_{B}}(b|a) $ for a fixed marginal distribution $ W_{\vec{e}_{A}}(A) $, however, it is not obvious that it is also convex in the $ \kappa $-$ \beta $-plane. We now show, that the mutual information~$ I $ is a convex function in the $ \kappa $-$ \beta $-plane, that is 
\begin{equation}\label{eq:convexity-ineq}
I(\lambda \kappa_{1} + (1-\lambda)\kappa_{2}, \lambda \beta_{1} + (1-\lambda)\beta_{2}) \leq \lambda I(\kappa_{1},\beta_{1}) + (1-\lambda)I(\kappa_{2},\beta_{2})
\end{equation}
for every $ \lambda $ with $ 0\leq \lambda \leq 1$.

We prove the relation, \cref{eq:convexity-ineq}, to be true, by starting from the right hand side of the inequality. By definition, we find 
\begin{align}
&\lambda I(\kappa_{1},\beta_{1}) + (1-\lambda)I(\kappa_{2},\beta_{2})\nonumber \\ &= \frac{\lambda}{4}\sum_{a,b} \left(1 + (-1)^{b}\beta_{1} + (-1)^{a+b} \kappa_{1}\right)\log_{2}\left(1+\frac{(-1)^{a+b} \kappa_{1}}{1+(-1)^{b}\ \beta_{1} }\right) \nonumber \\ &\quad+\frac{1-\lambda}{4}\sum_{a,b} \left(1 + (-1)^{b}\beta_{2} + (-1)^{a+b} \kappa_{2}\right)\log_{2}\left(1+\frac{(-1)^{a+b} \kappa_{2}}{1+(-1)^{b} \beta_{2} }\right). \label{eq:appKonvex1}
\end{align}
By introducing the abbreviations
\begin{equation}\label{key}
x_{1} \equiv \frac{\lambda}{4}\left(1 + (-1)^{b}\beta_{1} + (-1)^{a+b} \kappa_{1}\right)
\end{equation}
and 
\begin{equation}\label{key}
x_{2} \equiv \frac{1-\lambda}{4}\left(1 + (-1)^{b}\beta_{2} + (-1)^{a+b} \kappa_{2}\right),
\end{equation}
as well as
\begin{equation}\label{key}
y_{1} \equiv \frac{\lambda}{4}\left(1 + (-1)^{b}\beta_{1}\right)
\end{equation}
and 
\begin{equation}\label{key}
y_{2} \equiv \frac{1-\lambda}{4}\left(1 + (-1)^{b}\beta_{2}\right),
\end{equation}
we can simplify \cref{eq:appKonvex1} and find
\begin{align}
\lambda I(\kappa_{1},\beta_{1}) + (1-\lambda)I(\kappa_{2},\beta_{2}) = \sum_{a,b}\sum_{i=1}^{2}x_{i}\log_{2}\left(\frac{x_{i}}{y_{i}}\right).
\end{align}
According to the log sum inequality \cite{ThomasM.Cover2006} we have
\begin{equation}\label{key}
\sum_{i=1}^{2}x_{i}\log_{2}\left(\frac{x_{i}}{y_{i}}\right) \geq x \log_{2}\left(\frac{x}{y}\right)
\end{equation}
with $ x = x_{1} + x_{2} $ and $ y = y_{1} + y_{2} $. 

Hence, we find
\begin{equation}\label{key}
\lambda I(\kappa_{1},\beta_{1}) + (1-\lambda)I(\kappa_{2},\beta_{2}) \geq \sum_{a,b}x\log_{2}\left(\frac{x}{y}\right).
\end{equation}
By explicitly calculating $ x $ and $ y $ and comparing it with the definition of the mutual information we find 
\begin{equation}\label{key}
\sum_{a,b}x\log_{2}\left(\frac{x}{y}\right) = I(\lambda \kappa_{1} + (1-\lambda)\kappa_{2}, \lambda \beta_{1} + (1-\lambda)\beta_{2})
\end{equation}
Hence, we finally have 
\begin{equation}\label{key}
\lambda I(\kappa_{1},\beta_{1}) + (1-\lambda)I(\kappa_{2},\beta_{2}) \geq I(\lambda \kappa_{1} + (1-\lambda)\kappa_{2}, \lambda \beta_{1} + (1-\lambda)\beta_{2}) ,
\end{equation}
proofing the convexity of the mutual information.

\subsection{Extrema}
Due to the convexity of the mutual information, the maximum of the mutual information lies on the boundary of the ellipse. Hence, it is sufficient to restrict ourselves to the constraint
\begin{equation}\label{key}
\left(\frac{\kappa}{\mathcal{C}}\right)^{2} + \left(\frac{\beta}{\sqrt{1-\mathcal{C}^{2}}}\right)^{2} = 1,
\end{equation}
which is an equality instead of an inequality. 

We can parametrize the ellipse by an angle $ \varphi $, such that we have
\begin{equation}\label{eq:appB_k-ito-phi}
\kappa(\varphi) = \mathcal{C}\cos\varphi
\end{equation}
and
\begin{equation}\label{eq:appB_s-ito-phi}
\beta(\varphi) = \sqrt{1-\mathcal{C}^{2}}\sin\varphi.
\end{equation}
Inserting these two equations back into \cref{eq:appB_I-ks}, the mutual information becomes a function only dependent on a single parameter $ \varphi $. In order to maximize this function, we calculate the derivative with respect to $ \varphi $:
\begin{equation}\label{eq:appB_difIphi}
\dif{I(\varphi)}{\varphi} = \pdif{I(\kappa,\beta)}{\kappa}\dif{\kappa}{\varphi} + \pdif{I(\kappa,\beta)}{\beta}\dif{\beta}{\varphi}.
\end{equation}

First, from \cref{eq:appB_k-ito-phi,eq:appB_s-ito-phi}, we obtain the derivatives 
\begin{equation}\label{eq:appB_difk-ito-s}
\dif{\kappa}{\varphi} = - \mathcal{C}\sin\varphi = -  \frac{\mathcal{C}}{\sqrt{1-\mathcal{C}^2}}\beta(\varphi)
\end{equation}
and
\begin{equation}\label{eq:appB_difs-ito-k}
\dif{\beta}{\varphi} = \sqrt{1- \mathcal{C}^2}\cos\varphi =   \frac{\sqrt{1-\mathcal{C}^2}}{\mathcal{C}}\kappa(\varphi).
\end{equation}
We will now calculate the partial derivatives of the mutual information with respect to $ \kappa $ and $ \beta $. For the derivative with respect to $ \kappa $, we find
\begin{equation}\label{key}
\pdif{I}{\kappa} = \frac{1}{4}\sum_{a,b}(-1)^{a+b}\log_{2}\left(1 + (-1)^{a+b}\frac{\kappa}{1+(-1)^{b}\beta}\right) + \frac{1}{4 \ln 2}\sum_{a,b}(-1)^{a+b}.
\end{equation}
The second sum vanishes due to symmetry, such that we are left with
\begin{equation}\label{eq:appB_pdifIk}
\pdif{I}{\kappa} = \frac{1}{4}\sum_{a,b}(-1)^{a+b}\log_{2}\left(1 + (-1)^{a+b}\frac{\kappa}{1+(-1)^{b}\beta}\right),
\end{equation}
which is in general non-vanishing.

The derivative with respect to $ \beta $ is given by
\begin{equation}\label{key}
\pdif{I}{\beta} = \frac{1}{4}\sum_{a,b}(-1)^{b}\log_{2}\left(1 + (-1)^{a+b}\frac{\kappa}{1+(-1)^{b}\beta}\right) + \frac{1}{4 \ln 2}\sum_{a,b}(-1)^{a}\frac{\kappa}{1+(-1)^{b}\beta}.
\end{equation}
The second sum vanishes again due to symmetry relations, and we find
\begin{equation}\label{key}
\pdif{I}{\beta} = \frac{1}{4}\sum_{a,b}(-1)^{b}\log_{2}\left(1 + (-1)^{a+b}\frac{\kappa}{1+(-1)^{b}\beta}\right).
\end{equation}
When we insert this result together with \cref{eq:appB_pdifIk} into \cref{eq:appB_difIphi}, we obtain
\begin{align}\label{key}
\dif{I(\varphi)}{\varphi} = \frac{1}{4}\sum_{a,b}&\left((-1)^{b}\frac{\sqrt{1-\mathcal{C}^2}}{\mathcal{C}}\kappa(\varphi) - (-1)^{a+b}\frac{\mathcal{C}}{\sqrt{1-\mathcal{C}^2}}\beta(\varphi) \right)\nonumber \\ & \times \log_{2}\left(1 + (-1)^{a+b}\frac{\kappa}{1+(-1)^{b}\beta}\right)
\end{align}
This derivative has roots at $ \beta=0 $ and $ \kappa=0 $. Unfortunately, it is not obvious from an analytical point of view that those are the only two extrema. However, numerical simulations show, that these are the only ones.

For $ \kappa=0 $ it follows from \cref{eq:appB_I-ks}, that the mutual information vanishes for every value of $ \beta $. Since the mutual information cannot be negative, $ \kappa=0 $ represents a minimum of the mutual information. 

\subsection{Maximum}
We finally proof that $ \beta=0 $ is indeed a maximum of the mutual entropy. In order to do so, we take a look at the second order derivative
\begin{align}\label{key}
\dif{^2I(\varphi)}{\varphi^2} = & \pdif{^2I(\kappa,\beta)}{\kappa^2}\left(\dif{\kappa}{\varphi}\right)^2 +
2 \pdif{^2 I(\kappa,\beta)}{\beta\partial \kappa}\dif{\beta}{\varphi} \dif{\kappa}{\varphi}
+ \pdif{^2 I(\kappa,\beta)}{\beta^2}\left(\dif{\beta}{\varphi}\right)^2 \nonumber \\ &+
\pdif{I(\kappa,\beta)}{\kappa}\dif{^2 \kappa}{\varphi^2}  + \pdif{I(\kappa,\beta)}{\beta}\dif{^2 \beta}{\varphi^2}, 
\end{align}
which, in the case of $ \beta=0 $, simplifies to
\begin{equation}\label{key}
\left. \dif{^2I(\varphi)}{\varphi^2}\right|_{\beta=0} = 
\left. \pdif{I(\kappa,\beta)}{\kappa}\dif{^2 \kappa}{\varphi^2} \right|_{\beta=0}+ \left. \pdif{^2 I(\kappa,\beta)}{\beta^2}\left(\dif{\beta}{\varphi}\right)^2\right|_{\beta=0} .
\end{equation}
Calculating both terms explicitly, we find 
\begin{equation}\label{key}
\left. \pdif{I(\kappa,\beta)}{\kappa}\dif{^2 \kappa}{\varphi^2} \right|_{\beta=0} = - \frac{\mathcal{C}}{2}\left(\log_{2}\left(1+\mathcal{C}\right) - \log_{2}\left(1-\mathcal{C}\right) \right)
\end{equation}
as well as 
\begin{equation}\label{key}
\left. \pdif{^2 I(\kappa,\beta)}{\beta^2}\left(\dif{\beta}{\varphi}\right)^2\right|_{\beta=0} = \frac{\mathcal{C}^2}{\ln 2}.
\end{equation}
Hence, we arrive at
\begin{equation}\label{eq:appB_sdIphis0}
\left. \dif{^2I(\varphi)}{\varphi^2}\right|_{\beta=0} = \frac{\mathcal{C}^2}{\ln 2} - \frac{\mathcal{C}}{2}\left(\log_{2}\left(1+\mathcal{C}\right) - \log_{2}\left(1-\mathcal{C}\right) \right) .
\end{equation}
Since the values of $ \mathcal{C} $ are restricted to the interval $ 0 < \mathcal{C} < 1 $, we can evaluate the logarithms with help of the series representation
\begin{equation}\label{key}
\ln(1+ x) = \sum_{n=1}^{\infty} (-1)^{n+1}\frac{x^n}{n}
\end{equation}
valid for $ \abs{x} < 1 $, and the relation 
\begin{equation}\label{key}
\log_{2}x = \frac{\ln x}{\ln2}
\end{equation}
for converting the binary to the natural logarithm leads us to the identity
\begin{equation}\label{key}
\log_{2}\left(1+\mathcal{C}\right) - \log_{2}\left(1-\mathcal{C}\right) = \frac{2}{\ln2} \sum_{n=0}^{\infty} \frac{\mathcal{C}^{2n+1}}{2n+1}
\end{equation}
or
\begin{equation}\label{key}
\log_{2}\left(1+\mathcal{C}\right) - \log_{2}\left(1-\mathcal{C}\right)= \frac{2\mathcal{C}}{\ln2} + \frac{2}{\ln2} \sum_{n=1}^{\infty} \frac{\mathcal{C}^{2n+1}}{2n+1}.
\end{equation}

When we insert this relation into \cref{eq:appB_sdIphis0}, we find
\begin{equation}\label{eq:appB_sdIphis0-final}
\left. \dif{^2I(\varphi)}{\varphi^2}\right|_{\beta=0} =  - \frac{\mathcal{C}}{\ln2} \sum_{n=1}^{\infty} \frac{\mathcal{C}^{2n+1}}{2n+1} \leq 0,
\end{equation}
with equality if and only if $ \mathcal{C} = 0 $. Thus, the extremum $ \beta=0 $ corresponds to a maximum.

\section{Random measurements of the user}
\label{sec:rm}
In \cref{sec:wcs} we have considered the case in which the same projective measurement direction was chosen in each subsystem and for each experimental run. However, in general both the user and the attacker are  not restricted to a specific measurement direction but can select in each measurement a different one. 
In this Appendix, we discuss the special case in which the user is able to choose between \textit{two} distinct measurement directions at random, while we assume that the attacker stays with \textit{one}. 

This procedure is not necessarily the best approach for the attacker to pursue in order to maximize his information on the user's bit, but a realistic one if the attacker has neither the possibility to know the user's specific choice each time, or if he can only act passively, that is he cannot control the measurement on the environment.  

If, on the other hand, the attacker knew the measurement strategy, he could also perform measurements in two directions, correlated to the user's measurements. In this case the user's advantage is lost, since it reduces to the case of a single measurement direction in both $A$ and $B$, discussed in \cref{sec:singleprojmeas}. 

\subsection{Joint probabilities}

We now consider the scenario in which the user randomly chooses with equal probability from the two measurement directions $\vec{e}_{A}^{(1)}$ and $\vec{e}_{A}^{(2)}$ which are both perpendicular to the Bloch vector $ \vec{a}_{A} $, but differ by an angle $ \gamma $ with $0\leq \gamma \leq \pi$.

Here, the constraint of the vectors being perpendicular to the Bloch vector, is again made in order to obtain uniformly distributed bits $a$, that is 
\begin{equation}\label{eq:key-random-star}
W_{\vec{e}_{A}^{(1)}}(a)=W_{\vec{e}_{A}^{(2)}}(a)= \frac{1}{2}
\end{equation}
following from \cref{eq:prAa-2}. In contrast, the attacker uses a single measurement direction $\vec{e}_{B}$.

The joint probability 
\begin{align}
W_{\lbrace\vec{e}_{A}^{(j)}\rbrace,\vec{e}_{B}}(a,b) 
=\frac{1}{2}\left(W_{\vec{e}_{A}^{(1)},\vec{e}_{B}}(a,b) + W_{\vec{e}_{A}^{(2)},\vec{e}_{B}}(a,b) \right),\label{eq:prAB-ito-angles_2b0} 
\end{align}
is the average value of the probabilities $ W_{\vec{e}_{A}^{(1)},\vec{e}_{B}} $ and $ W_{\vec{e}_{A}^{(2)},\vec{e}_{B}} $, which are given by \cref{eq:joint-prob_explicit}, of the individual measurement directions, since both measurement directions $ \vec{e}_{A}^{(1)} $ and $ \vec{e}_{A}^{(2)} $ are independent of each other and occur with the same probability. 

We write \cref{eq:prAB-ito-angles_2b0} in the form
\begin{align}
W_{\lbrace\vec{e}_{A}^{(j)}\rbrace,\vec{e}_{B}}(a,b) 
=&  \frac{1}{4} \left(1 + (-1)^{b}\beta +(-1)^{a+b} \kappa_{\text{eff}} \right)\label{eq:prAB-ito-angles_2b} 
\end{align}
with a new effective correlation parameter
\begin{equation}\label{key}
\kappa_{\text{eff}} = \left( \frac{\vec{e}_{A}^{(1)}+\vec{e}_{A}^{(2)}}{2}\right)^\top \tilde{K}\vec{e}_{B}
\end{equation}
When we define the unit vector 
\begin{equation}\label{eq:def-eplus}
\overline{\vec{e}}_{A}\equiv \frac{\vec{e}_{A}^{(1)}+\vec{e}_{A}^{(2)}}{\abs{\vec{e}_{A}^{(1)}+\vec{e}_{A}^{(2)}}}, 
\end{equation}
which is again perpendicular to the Bloch vector $ \vec{a}_{A} $, we obtain
\begin{align}
\kappa_{\text{eff}}=\cos\left(\frac{\gamma}{2}\right) \overline{\vec{e}}_{A}K\vec{e}_{B} ,\label{eq:prAB-ito-angles_2b-final} 
\end{align}
since we have
\begin{equation}\label{key}
\abs{\vec{e}_{A}^{(1)}+\vec{e}_{A}^{(2)}}= \sqrt{2(1+\cos\gamma)}=2\cos\left(\frac{\gamma}{2}\right).
\end{equation}

Apart from the constant factor $ \cos(\gamma/2) $ the correlation parameter $ \kappa_{\text{eff}} $, \cref{eq:prAB-ito-angles_2b-final}, is the same as the correlation parameter $ \kappa $, \cref{eq:def_alphabetakappa}, for the case of single measurement. 

By using \cref{eq:def_K}, together with $ \overline{\vec{e}}_{A,z}=0 $, we find
\begin{equation}\label{eq:kappa-eff-explicit}
\kappa_{\text{eff}}= \mathcal{C}_{\text{eff}} \overline{\vec{e}}_{A,x}\vec{e}_{B,x} - \mathcal{C}_{\text{eff}} \overline{\vec{e}}_{A,y}\vec{e}_{B,y},
\end{equation}
with the effective correlation
\begin{equation}\label{eq:Imax_random-meas2}
\mathcal{C}_{\text{eff}} \equiv \mathcal{C}\cos\left( \frac{\gamma}{2}\right).
\end{equation}

\subsection{Discussion and caviat}
By comparing \cref{eq:kappa-eff-explicit} with \cref{eq:kappa-explicit} for the case of $ \vec{e}_{A,z}=0 $, we see that they only differ by in their concurrence. 

Hence, the maximal mutual information still has the form of \cref{eq:Imax_for_chi=pihalf}, with the concurrence $\mathcal{C}$ being replaced by $ \mathcal{C}_{\text{eff}} $.


The case $\gamma =0$, that is when both measurements coincide with another, reduces to the one of a single measurement direction, discussed in \cref{sec:wcs}. However, for $\gamma > 0$, we have $\cos(\gamma/2) < 1$, and thus the maximal mutual information is decreased compared to a single measurement direction. Indeed, by choosing $\gamma = \pi$, the maximal achievable mutual information is reduced to $ I_{\max}=0$, independent of the concurrence of the state. 

In this scenario the user randomly chooses orthogonal measurement directions. Hence, the randomness originates from the fact that he randomly assigns different bit values to the same measurement result. As a consequence, the user would need another QRNG to create this randomness, in this way he puts turtles on top of turtles.

%
%



\section*{References}

\begin{thebibliography}{10}
\bibitem{Cutshall2020}
	Cutshall J~M, McPhee T and Beck M State preparation and measurement tomography
	via unitary transformations (To be published)
\end{thebibliography}
\providecommand{\newblock}{}

\end{document}